\documentclass[12pt,preprint]{aastex}
  \newcommand{\ms}{\,\rm m\,s^{-1}}
  
  \newcommand{\kms}{\,\rm km^2\,s^{-1}}
  \newcommand{\dgr}{^{\circ}}

\begin{document}

\title{Solar cycle prediction using precursors and flux transport models}
\author{R. Cameron and M. Sch\"ussler}
\affil{Max Planck Institut f\"ur Sonnensystemforschung, 
       Max-Planck-Str.~2,
       37191 Katlenburg-Lindau, Germany}

\email{cameron@mps.mpg.de,schuessler@mps.mpg.de}

\begin{abstract}
%Version: \today
We study the origin of the predictive skill of some methods to
forecast the strength of solar activity cycles. A simple flux
transport model for the azimuthally averaged radial magnetic field at
the solar surface is used, which contains a source term describing the
emergence of new flux based on observational sunspot data. We consider
the magnetic flux diffusing over the equator as a predictor, since this
quantity is directly related to the global dipole field from which a
Babcock-Leighton dynamo generates the toroidal field for the next
activity cycle. If the source is represented schematically by a narrow
activity belt drifting with constant speed over a fixed range of
latitudes between activity minima, our predictor shows considerable
predictive skill with correlation coefficients up to 0.95 for past
cycles. However, the predictive skill is completely lost when the
actually observed emergence latitudes are used. This result originates
from the fact that the precursor amplitude is determined by the sunspot
activity a few years before solar minimum. Since stronger cycles tend to
rise faster to their maximum activity (known as the Waldmeier effect),
the temporal overlapping of cycles leads to a shift of the minimum
epochs that depends on the strength of the following cycle. This
information is picked up by precursor methods and also by our flux
transport model with a schematic source. Therefore, their predictive
skill does not require a memory, i.e., a physical connection between the
surface manifestations of subsequent activity cycles.
\end{abstract}


\section{Introduction}
\label{sec:intro}

%As the activity minimum of the current solar cycle approaches,  
%predictions of the strength of the next cycle ...

A prediction of the strength of future solar cycles is of interest
because a) expected future levels of solar activity have implications
for space missions and for foreseeing potential hazards due to `space
weather', and b) it can be used to test theoretical models of the solar
cycle, if the prediction is based upon a physical approach.  Existing
attempts to predict solar activity levels can be broadly divided into
two groups: 1) extrapolation models, deriving a prediction from a purely
mathematical analysis of the past records of solar activity, for
instance by harmonic analysis \citep[e.g.,][]{Rigozo:etal:2001,
Echer:etal:2004} or by using concepts from nonlinear dynamics
\citep[e.g.,][]{Sello:2001}, and 2) precursor models based upon
correlations between certain measured quantities in the declining phase
of a cycle and the strength of the next cycle
\citep[e.g.,][]{Hathaway:etal:1999, Schatten:2003}. While the overall
success of these models in predicting the {\em future} has been rather
limited (e.g., see Figure~14.2 in Wilson 1994 and Figure~6 of Lantos \&
Richard 1998), a number of methods have demonstrated considerable skill
in `predicting' the strength of the past solar cycles
\citep[e.g.][]{Hathaway:etal:1999}. In particular, some methods based on
the level of geomagnetic variations (as measured, e.g., by the $aa$ or
$A_p$ indices) a few years before and around sunspot minimum provide
high correlation coefficients of up to 0.97 between the respective
predictor and past maxima of the sunspot record
\citep[e.g.,][]{Legrand:Simon:1981,Layden:etal:1991,Thompson:1993,
Lantos:Richard:1998, Hathaway:etal:1999}

Since the recurrent geomagnetic variations in the late phases of a cycle
are presumably dominated by the fast solar wind streams from the
equatorward expanding polar coronal holes of the Sun, some relation to
the strength of the polar magnetic field during that phase has been
suggested.  As noted by, e.g., \citet{Schatten:etal:1978} and
\citep{Layden:etal:1991}, such a relationship would be generally
consistent with the Babcock-Leighton type of dynamo models for the solar
cycle \citep{Babcock:1961, Leighton:1969, Giovanelli:1985, Wang:Sheeley:1991,
Choudhuri:etal:1995, Durney:1995, Durney:1996, Dikpati:Charbonneau:1999,
Kueker:etal:2001}. Such models assume that the differential rotation
winds up a dipolar global poloidal field which is thought to dominate
the large-scale surface field distribution around solar minimum and
creates sub-surface toroidal fields, whose later eruption causes the
magnetic activity of the following cycle. The larger the global dipole
moment is, the stronger the toroidal field becomes and, presumably, the
stronger the next cycle will be. In the course of the cycle, the global
dipole field is reversed and an opposite-polarity field built up, owing
to the preferential poleward drift of the follower polarity parts of
solar active regions and the diffusion and cancellation of
preceding-polarity flux over the equator. This occurs because there is a
(statistically) systematic tilt angle of the bipolar magnetic regions
with respect to the equator (Joy's law). The important point is that the
mechanism which provides the poloidal field for the next cycle operates
in the near-surface layers and thus is directly accessible to
observation while the old cycle is still ongoing. If quantitatively
understood this could provide a physical basis for prediction and,
at the same time, a testbed for dynamo models.

The first attempt to actually use a theoretical dynamo model for
predicting future solar activity levels was presented recently by
\citet{Dikpati:etal:2006} and \citet{Dikpati:Gilman:2006}, in the
following referred to as the DDG model. They considered a
Babcock-Leighton type flux-transport dynamo with solar-like differential
rotation, advection of the magnetic field by a prescribed meridional
circulation pattern (poleward near the surface, equatorward in the lower
convection zone) and two sources for the poloidal field: one near the
bottom of the convection zone and one resulting from active-region tilt
near the surface. Parametrizing the near-surface source with the
observed record of sunspot areas since 1874 and making simple
assumptions concerning the tilt angle and the equatorward migration of
the source latitude (corresponding to the butterfly diagram), they took
the amount of flux of the low-latitude toroidal field generated by the
model as a measure of predicted solar activity. Their model is able to
reproduce the variation of the amplitudes of cycles 16--23 remarkably
well, with correlation coefficients reaching up to nearly 0.99,
depending on the model parameters.

The success of the DGG model suggests that it may have captured the
essential ingredients of the solar dynamo mechanism and that the dynamo
is of the Babcock-Leighton flux-transport type. This would be an
enormous advance in our understanding and, therefore, the evidence must
be evaluated thoroughly and critically. We have therefore set out to
answer the following questions: 1) Which are the key physical features
that underly the predictive skill of the the DGG model? 2) How much of
the predictive skill is already contained in the surface transport part?
3) Could the predictive skill be affected by correlations in the sunspot
record (used as input data for the model) that are not specific to any
particular dynamo model?

In order approach answers to the first two questions, we have carried
out an exploratory study with a simple surface flux transport
model. Two-dimensional versions of such models have successfully
reproduced the observed evolution of the solar surface field
\citep[e.g.,][]{Wang:etal:1989, Wang:etal:1989b,
Ballegooijen:etal:1998a, Schrijver:2001, Durrant:etal:2004,
Baumann:etal:2004}. For simplicity (and consistent with the surface part
of the DGG model) we have considered azimuthally averaged quantities in
a one-dimensional flux transport model and determined the evolution of
the latitude-dependent surface field on the basis of observational input
data (sunspot groups or sunspot numbers). If the generation of the
poloidal field for the next cycle actually takes place near the surface
in the fashion exhibited by the Babcock-Leighton approach, we should be
able to find relevant predictors from such results. We have therefore
compared quantities related to the reversal and built-up of the global
dipole field, particularly the amount of flux diffusing and cancelling
over the equator, with the activity of the following cycle.

In order to address the third question, we have also studied whether the
predictive skill of the precursor-type models (including our own and
possibly also DDG model) could be affected by the overlap of consecutive
cycles and the resulting shift of the epoch of activity minimum
depending on the strength and the rise time of the following cycle.

The paper is organized as follows. We describe the flux transport model
in Sect.~\ref{sec:ftm} and define the quantities considered as
predictors in Sect.~\ref{sec:predictors}. Results based upon
observational datasets for sunspot areas and sunspot numbers are shown
in Sect.~\ref{sec:results}. We first closely follow the procedure used
by DDG and compare with their results. Then we evaluate the sensitivity
of the results to parameter variations and to modifications of the model
assumptions concerning the input data, particularly the latitude
distribution of emerging flux.  In Sect.~\ref{sec:origin} we consider
series of synthetic cycles with random amplitudes in order to study the
effect of the amplitude-dependent shift of the activity minimum epoch on
prediction methods.  Sect.~\ref{sec:discussion} contains a discussion of
the results and gives our conclusions.

\section{A simple flux transport model}
\label{sec:ftm}

We model the evolution of the (azimuthally averaged) component of the
field normal to the solar surface, $B_r$, as a function of latitude,
$\lambda$, and time, $t$. We assume that the surface flux is passively
advected by a meridional flow, $v(\lambda)$, and by supergranulation,
the latter being described as a turbulent diffusion process with an
effective diffusivity $\eta$. In addition, we have a source term,
$S(\lambda,t)$, describing the emergence of new bipolar magnetic regions
at the surface, and an exponential decay term with a characteristic time
scale $\tau$, which roughly mimics 3-D effects
\citep{Schrijver:etal:2002, Baumann:etal:2006}. Accordingly, our simple
1-D surface transport equation is written as
\begin{eqnarray}
\frac{\partial B_r}{\partial t} &=& 
  \frac{1}{R_{\odot}\cos\lambda}\frac{\partial}{\partial \lambda} 
    \bigg[ v(\lambda) B_r \cos\lambda \bigg] +
     \nonumber\\ \noalign{\vspace{0.2cm}} 
  &+& \frac{\eta}{R_{\odot}^2\cos\lambda}
  \frac{\partial}{\partial \lambda} 
  \left( \sin\lambda\frac{\partial B_r}{\partial\lambda}\right) 
  - \frac{B_r}{\tau} + S(\lambda,t)\,.
\label{eq:transport}
\end{eqnarray}
For simplicity, we take $B_r$ to be antisymmetric with respect to
the equator $(\lambda=0$) and consider only the Northern hemisphere,
thus ignoring North-South asymmetries. 

We solve this 1D linear advection-diffusion problem by a fully implicit
finite difference scheme with 900 grid cells per hemisphere. Since the
bipolar regions described by the source term typically have a size of a
few degrees, all features occurring in the simulation are very well
resolved. In fact, tests with only 90 grid points give essentially the
same results.  The timestep of the numerical integration is set
according to the Courant-Friedrichs-Levy condition on the basis of the
maximum meridional flow speed. Values typically are of the order of a
day.

\section{Predictors}
\label{sec:predictors}

Which quantities provided by the surface flux transport model could be
considered as potential predictors for the strength of the next cycle?
In the framework of Babcock-Leighton dynamos, one is tempted to use the
maximum value of the reversed polar field built up after solar maximum
as an indicator for the strength of the poloidal field from which the
toroidal field of the next cycle is being generated by differential
rotation \citep{Schatten:etal:1978}. In fact,
\citep{Svalgaard:etal:2005} have reported a correlation between observed
polar fields and sunspot activity of the subsequent cycle, but regular
measurements of the polar fields exist only since about 1970 and still
are rather uncertain and noisy \citep[see,
e.g.,][]{Arge:etal:2002}. Furthermore, \citet{Layden:etal:1991} have
tried to extend the temporal basis by considering potential proxies for
the polar field such as the shape of the polar corona, coronal holes, or
polar faculae, but found no clear evidence of a `predictive' skill.

Even in the framework of the Babcock-Leighton scenario, it is not so
clear whether the maximum polar field strength necessarily represents
the relevant quantity that determines the strength of the next cycle. In
a flux transport dynamo, the amount of poloidal magnetic flux reaching
deep into the convection zone (the global dipole) is relevant for the
generation of the toroidal field emerging during the next cycle. The
actual polar surface field, on the other hand, contains also more
superficial flux which is still topologically connected with the
corresponding preceding-polarity flux on the same
hemisphere. Furthermore, high-latitude flux emergence could possibly
also contribute to the polar field \citep{Durrant:etal:2002}. The
reversal and built-up of the global dipole field relevant for a
flux-transport dynamo is determined by the amount of preceding-polarity
flux that diffuses across the equatorial plane and reconnects with the
opposite-polarity preceding flux on the other hemisphere. This is nicely
illustrated by Figure~10 of \citet{Dikpati:Gilman:2006}.  We therefore
consider the amount of flux crossing the equator (per unit time, or
time-integrated) as the potentially most relevant precursor for the
strength of the subsequent cycle in the framework of a Babcock-Leighton
dynamo. It can easily be calculated in surface flux transport
simulations and it also lends itself to observational determination
\citep{Durrant:etal:2004}. 
Note that this predictor differs from what is used by DGG, namely,
the low-latitude toroidal flux at the bottom of the convection zone,
which is not present in our model. However, in a linear Babcock-Leighton
type flux transport dynamo, the generated toroidal field is directly
related to the poloidal flux diffusing over the equator. Thus there is a
physical link between the two models, although we do not claim that our
model provides a direct test of the DGG model.

\section{Results of the flux transport model}
\label{sec:results}

We have considered two sets of simulation runs with our flux transport
model. In the first set, the results of which are presented in
Sect.~\ref{subsec:Dikpati}, we have followed as closely as possible the
procedures in the DDG model in order to study whether part
of the predictive skill of their model is already contained in the
surface flux transport. The second set of simulations addresses the
dependence of the predictive skill on the parameters of the flux
transport model as well as on the assumptions and procedures concerning
the emergence latitudes of tilted bipolar regions.

\subsection{Input model according to Dikpati et al.(2006)}
\label{subsec:Dikpati} 

The DDG model \citep{Dikpati:etal:2006} is described in some more
detail in \citet{Dikpati:Gilman:2006}. Excluding parameters of the
dynamo model which are of no concern here, the procedures can be
described as follows:
\begin{enumerate}
\item The surface source term for the poloidal magnetic field (more
  precisely, for the azimuthal component of the vector potential) is
  taken proportional to the sunspot areas given in the combined RGO/SOON
  observational record provided by D. Hathaway
  (NASA/MSFC)\footnote{http://solarscience.msfc.nasa.gov/greenwch.shtml}.
\item The individual solar cycles are stretched or
  compressed in time to fit with a mean cycle length of 10.75 yr.
\item The source term has a Gaussian latitudinal profile with a fixed
full width at half maximum (FWHM) of $6\dgr$, migrating equatorward
between $35\dgr$ and 5$\dgr$ between cycle minima.
\item The magnetic diffusivity in the near-surface layers has a value of
  about 300$\,\kms$.
\item The poleward meridional flow at the surface has the form
  $v(\lambda)=v_0\,\sin(2\lambda)$ with $v_0=14.5\ms$.
\end{enumerate}
We have adapted our surface transport model as far as possible to these
parameters and procedures. We describe the source term (in our case,
written for $B_r$) by two Gaussians of opposite sign with a FWHM of
6$\dgr$, centered at $\lambda_0(t)\pm 0.\dgr5$, where $\lambda_0(t)$
parametrizes the equatorward drift of the flux emergence latitude
between $35\dgr$ and $5\dgr$ as a linear function of cycle phase.  The
amplitude of the Gaussians as a function of time is taken to be
proportional to the monthly averages of the sunspot area from the
RGO/SOON dataset, smoothed by a 6-month boxcar running mean. For the
decay time, $\tau$, we use a value of 5.7 years \citep[roughly
corresponding to a volume diffusivity of about 100$\kms$,
see][]{Baumann:etal:2006}.

The epochs of the sunspot minima have been taken from the National
Geophysical Data Center
(NOOA/NGDC)\footnote{ftp://ftp.ngdc.noaa.gov/STP/SOLAR\_DATA/SUNSPOT\_NUMBERS/maxmin.new}.
Since cycle 23 is still ongoing, we have considered three possible
values for the epoch of the upcoming sunspot minimum, namely 2007.0,
2007.5 and 2008.0, respectively. Since the RGO/SOON data are available
only until 10/2005, we have substituted the missing data between 11/2005
and the assumed minimum epoch by data for the corresponding phase of
cycle 22.

\begin{figure}[ht!]
\centering
\resizebox{0.6\hsize}{!}{\includegraphics{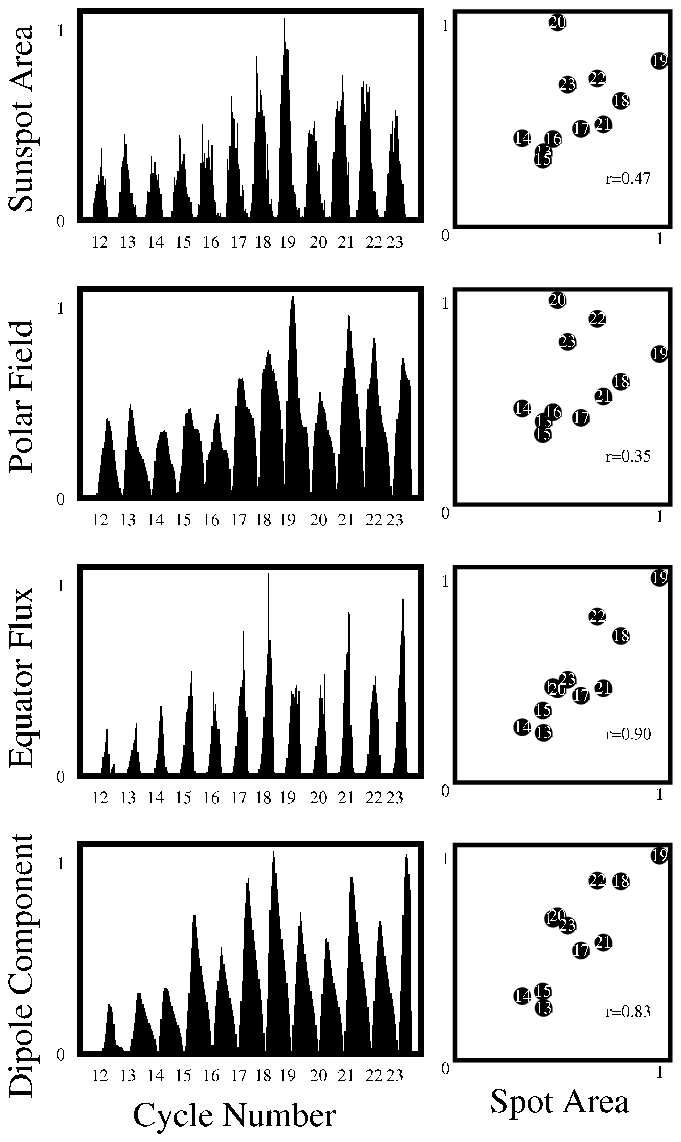}}
\caption{Time evolution of cycle-related quantities (left panels) and
corresponding correlation diagrams (right panels). Top row: RGO/SOON
sunspot areas and correlation diagram between subsequent maxima. Second
row: unsigned polar field resulting from the surface flux transport
model and correlation diagram of the maximum with the maximum spot area
of the subsequent cycle.  Third row: same for the magnetic flux
diffusing over the equator per unit time from the flux transport model.
Bottom row: same for the dipole component of the surface field from the
flux transport model. The flux transport model is adapted as far as
possible to the procedures and parameters used in the model of
\citet{Dikpati:etal:2006}. All quantities shown are normalized to their
respective global maxima. The numbers of the `predicted' cycles and the
correlation coefficients are indicated in the correlation diagrams.
Unless stated otherwise, all time-dependent quantities in this and the
following diagrams are binned in six-month totals after applying a
time-centered 6-month boxcar averaging.}
\label{fig:Dikpati_reference}
\end{figure}

Figure~\ref{fig:Dikpati_reference} shows the time evolution (for the
period covered by the RGO/SOON data) of various quantities and the
corresponding correlation diagrams. The top panels give the RGO/SOON
sunspot areas and the correlations between the maximum of a cycle with
that of the preceding cycle; obviously, the amplitude of the foregoing
cycle is a poor predictor (correlation coefficient $r=0.47$). The second
panel from the top shows the polar field as resulting from the flux
transport simulation. It peaks typically around sunspot
minimum and largely reflects the level of activity of the {\em ongoing}
cycle, a common result of surface flux transport models
\citep{Schrijver:etal:2002, Wang:etal:2002b, Wang:etal:2005,
Baumann:etal:2006}. Consequently, the predictive power is poor
($r=0.35$).

The third panel shows the amount of magnetic flux diffusing over the
equator per unit time, which we henceforth denote by $\Phi$, for
simplicity.  We consider this predictor to be related to the build-up of
the global poloidal field relevant for the dynamo. We see that this
quantity shows a reasonable predictive skill with $r=0.90$.  In
particular, the strong drop of activity from cycle 19 and 20 is
correctly anticipated. On the other hand, the prediction for the
relatively high cycle 21 is quite bad. Finally, the bottom panel gives
the dipole component of the simulated surface field, which mainly
results from the accumulated flux diffusing over the equator. It shows a
predictive power that is comparable to that of $\Phi$ ($r=0.83$).

As an alternative to correlating the maxima, we have also considered
time-integrated quantities, namely, cycle-integrated sunspot area and
magnetic flux diffused over the equator as well as polar field and
dipole component of the surface field integrated between polarity
reversals. All integrals have been determined for unstretched cycles,
i.e., keeping the actual cycle lengths.  The correlation coefficients of
the integrated sunspot area of the subsequent cycle with the integrated
flux over the equator ($r=0.92$) and the dipole component ($r=0.82$) are
similar to the corresponding values for the maxima. Interestingly, the
correlations coefficients for the integrated polar field and the sunspot
area of the preceding cycle are much larger: 0.79 and 0.78,
respectively. At least part of this increase comes from the better
`prediction' of cycle 20, whose lower amplitude is largely compensated
by a its longer duration in the integrals. This result indicates that
all such correlations should be considered with great caution because
they can be sensitive to the precise definition of the quantities
considered and their real significance is difficult to evaluate.

The results presented in Figure~\ref{fig:Dikpati_reference} show that a
significant part of the predictive skill contained in the DDG model is
covered by surface quantities resulting from a simple flux transport
model, although we do not quite reach such high correlation coefficients
when using their model parameters. The fact that the magnetic flux
diffusing over the equator appears to be a fairly good predictor would
be consistent with its putatively important role played in a
Babcock-Leighton dynamo. What is relevant for the amplitude $\Phi$?
Most importantly, it is the amount of flux emerging near to the equator
in the late phases of a cycle.  Since the distance to the equator is
small and since the poleward meridional flow accelerates away from the
equator, low-latitude bipolar magnetic regions contribute most to
$\Phi$.  Figure~\ref{fig:blowup} illustrates the importance of
low-latitude emergence during the declining phase of the cycle for the
predictor $\Phi$. Given are the observed sunspot area (solid) and the
flux diffusing over the equator, $\Phi$ (dashed), for cycles 18, 19, and
20. In addition, the dotted lines (with a separate axis on the right
hand side) show the linear progression of the centroid of the emergence
latitude assumed in the model. The maxima of $\Phi$ are shifted by about
3--4 years with respect to the preceding sunspot maxima and occur about
2--3 years before the subsequent minima.  Since the $\Phi$ curve
essentially represents a convolution of the sunspot area curve with the
latitude sensitivity of the flux diffusing over the equator (which is
heavily weighted towards the low latitudes), its amplitude is determined
by the sunspot area and the emergence latitude at a given time. The
sunspot area is much larger during the time interval when the emergence
latitude ranges between about $15\dgr$ and $5\dgr$ in the case of cycle
18 than for cycle 19. Consequently, the prediction for cycle 19 is much
higher than that for cycle 20. This result already indicates that the
predictor could be rather sensitive to the definition of the source
latitudes in the model.

A note concerning the prediction for cycle 24: the results shown in
Figure~\ref{fig:Dikpati_reference} are based on assuming 2007.0 for the
epoch of the upcoming sunspot minimum. This yields a predicted maximum
sunspot number of about 160 for cycle 24. However, a later minimum would
lead to significantly lower predictions: about 130 for a 2007.5 minimum,
and about 110 for a minimum at 2008.0. The first two values are
consistent with the DDG prediction of a cycle 24 amplitude exceeding
that of cycle 23 by 30--50\%, while the third case would predict an
activity similar to that of cycle 23. All these results are affected in
an unknown way by the substitution of data from cycle 22 for unknown
cycle 23 data. In any case, we see that the predictions rather
sensitively depend on the assumed epoch of the upcoming activity minimum
- the significance of this result will become clear when we consider the
origin of the predictive skill in Sect.~\ref{sec:origin}.

\subsection{Modifications of the procedure}
\label{subsec:modifications} 

We have seen that the amount of preceding-polarity flux diffusing over
the equator appears to represent a reasonably good predictor for the
next cycle. In this section we will evaluate a) how strongly this result
depends on parameter choices in our flux transport model and b) whether
the predictive skill can be improved by using more of the observational
information contained in the available sunspot data. It turns out that
the correlation coefficients for the maxima of the polar field are
always small, while the results for the dipole component of the surface
field and for the flux crossing the equator, $\Phi$, are similar to each
other in most cases. Therefore, in what follows we restrict ourselves to
showing results for the latter quantity.

\subsubsection*{Dependence on model parameters}
\clearpage
\begin{deluxetable}{cl|c|ccc }
  \tabletypesize{\small}
  \tablewidth{0pt}
  \tablenum{1}
  \tablecolumns{6}
  \tablecaption{Correlation coefficients 
    between the predictor, $\Phi$, and the strength of the
    following activity cycle \label{tab:ccs}}
  \tablehead{
    \colhead{\ \ } & \colhead{variable} & \colhead{reference} &
    \colhead{$a$}  & \colhead{$b$} & \colhead{$c$}}
  \startdata
 1. & meridional flow velocity, $v_0\,(\ms)$   
    & 14.5     & 10.     & 20.     &          \\
  & & {\it 0.90}   & {\it 0.88}  & {\it 0.85}  &          \\[6pt]
 2. & meridional flow profile 
    & DDG      & SB2006  &         &          \\
  & & {\it 0.90}    & {\it 0.82}  &         &          \\[6pt]
 3. & magnetic diffusivity, $\eta\,(\kms)$ 
    & 300.     & 100.    & 200.    & 600.     \\
  & & {\it 0.90}    & {\it 0.70}   & {\it 0.79}   & {\it 0.76}  \\[6pt]
 4. & decay time, $\tau$ (yr) 
    & 5.6     & 1.7    & 27.8   &          \\
  & & {\it 0.90}    & {\it 0.89}   & {\it 0.90}  &          \\[6pt]
 5. & tilt factor in source amplitude
    & none     & $\sin\lambda$  &  &          \\
  & & {\it 0.90}    & {\it 0.88}  &         &          \\[6pt]
 6. & Gaussian width, FWHM (deg)   
    & 6.       & 3.      & 12.     &          \\
  & & {\it 0.90}    & {\it 0.85}   & {\it 0.66}  &          \\[6pt]
 7. & Latitude shift of Gaussians (deg)
    & 1.       & 2.      & 3.      & 4.       \\
  & & {\it 0.90}    & {\it 0.90}   & {\it 0.90}   & {\it 0.91}   \\[6pt]
 8. & Latitude range of emerging flux (deg)
    & 5--35    & 1--31   & 5--25   & 5--40    \\
  & & {\it 0.90}    &  {\it 0.80}  &  {\it 0.74}  &  {\it 0.82}  \\[6pt]
 9. & stretching of cycles 
    & yes      & no      &         &          \\
  & & {\it 0.90}    & {\it 0.91}  &         &          \\[6pt]
    \enddata
\tablecomments{The first column indicates the variable or procedure of
  the model being varied. The subsequent columns show the results for
  the reference case (Sect.~\ref{subsec:Dikpati}) and the variations of
  the model. In each case, the upper number gives the value of the
  parameter under consideration and the lower number (in italics) the gives
  corresponding correlation coefficient.}
\end{deluxetable}
\clearpage
As a first step, we consider the effect of parameter variations in our
model on the correlation coefficient between the maximum of $\Phi$, the
amount of flux diffusing over the equator, and the activity maximum of
the subsequent cycle.  The basic procedure for including the emerging
flux as a source in the model is the same as in the reference case
discussed in the preceding section. We have varied the following
parameters:
\begin{enumerate}
\item {\em Peak meridional velocity, $v_0$.} 
\item {\em Latitude profile of the meridional flow, $v(\lambda)$.} As an
      alternative to the profile used by DDG, we consider a
      semi-empirical profile adapted to helioseismic measurements,
      $v(\lambda)=1.6\sin(2\lambda)\,\exp[\pi(1-\vert\lambda\vert/90\dgr)]$,
      in units of $\ms$ \citep{Schuessler:Baumann:2006}.
\item {\em Turbulent magnetic diffusivity, $\eta$.} The value of $600\kms$
      favored by surface flux-transport models
      \citep[e.g.,][]{Durrant:Wilson:2003, Baumann:etal:2004} is a
      factor of 2 larger than the surface value considered by DDG. We
      have also considered smaller values of $\eta$. 
\item {\em Decay time, $\tau$.} The results of \citet{Schrijver:etal:2002}
      and \citet{Baumann:etal:2006} favor a value between 5 and 10
      years. 
\item {\em Tilt angle factor in the source amplitude.} DDG have taken
      the amplitude of their source term proportional to the sunspot
      area as a proxy of the amount of emerging flux, thus implicitly
      assuming that the average tilt angle of the emerging bipolar
      regions is independent of latitude. In reality, the average tilt
      angle is roughly proportional to $\sin\lambda$
      \citep[e.g.,][]{Howard:1991}.  This can be included into the model
      by multiplying the source term by $\sin\lambda$.  The linearity of
      the flux transport model permits us to ignore the factor of
      proportionality since it represents only a scaling factor that
      does not affect the correlation results. 
\item {\em FWHM of the Gaussians in the source term.} DDG have assumed a
      fixed latitudinal width of $6\dgr$ for their source term,
      corresponding to a constant width of the activity belt in the
      course of the cycle.
\item {\em Latitude shift between the Gaussians.} As long as this shift is
      small compared to the width of the Gaussian, the sum of the two
      Gaussians (of opposite sign) used in our source term is
      proportional to the derivative of the Gaussian with respect to
      latitude and the width of the resulting source term is only determined
      by the FWHM of the individual Gaussians. The effect of moving away
      from this limit can be evaluated by increasing the shift from
      $1\dgr$ (reference case) to larger values.
\item {\em Range of emergence latitudes.} The procedure of DDG assumes that
      the source progresses through a fixed latitude interval in the
      course of each cycle. The boundaries of this interval are
      adjustable parameters. 
\item {\em Stretching of cycles.} In order to keep their dynamo model in phase
      with the source term derived from the actual sunspot data, DDG had
      to stretch (or compress) the lengths of the individual cycles to a
      fixed value of 10.75 years. Our simple model does not require such
      stretching, so that the actual cycle length can be used.
\end{enumerate}
Table~\ref{tab:ccs} gives the correlation coefficients between the
maximum of $\Phi$ and the maximum sunspot area of the subsequent cycle
for various parameter changes, following the sequence of the preceding
list. The correlation coefficients are printed in italic numbers in the
second row for each case. As reference we take the case presented in
Sect.~\ref{subsec:Dikpati}, which follows closely the DDG procedures.
We see that the most sensitive parameters are the magnetic diffusivity
and the latitudinal width of the source term, which is not surprising in
view of the importance of these parameters with respect to our
predictor, $\Phi$. The corresponding correlation coefficients based on
the cycle-integrated quantities are very similar to those shown in
Table~\ref{tab:ccs} and thus need not to be discussed any further.
Altogether, it is clear from the table that, as long as we keep the
basic procedures used by DDG, the correlation coefficients are not
strongly dependent on the parameters, with values around $r=0.90$ being
common. We can easily obtain even higher values by performing some fine
tuning: for instance, for $\eta=200\kms$ and a latitude range of
$5\dgr$--$25\dgr$ for the source term, we find a correlation coefficient
of 0.95 between the maxima of $\Phi$ and the sunspot area. However, such
a procedure is dubious for obvious reasons and we have not tried to
further `optimize' the values.

\subsubsection*{Dependence on input data set}

The cross-calibration between the RGO and the SOON data to obtain a
consistent data set spanning the whole time period from 1874 on is not
trivial. \citet{Balmaceda:etal:2005} have recently considered an
additional sunspot data set, the ``Russian books'', compiled from data
obtained at observatories in the former USSR.  Applying a careful
cross-calibration, they have bridged the RGO and the SOON data by parts
of this data set and thus obtained a consistent record of sunspot
areas. In order to evaluate the dependence of the predictive skill on
the input data set, we have taken these data as input for our flux
transport model. In addition, we also have used the monthly sunspot
numbers as provided by the Solar Influences Data Analysis Center
(SIDC)\footnote{http://sidc.oma.ce} at the Royal Observatory of
Belgium. Figure~\ref{fig:Dikpati_datasets} shows the quantity $\Phi$
obtained with the surface flux transport model using the three data sets
for cycles 12-23. All other parameters are the same as those in
Sect.~\ref{subsec:Dikpati}. As apparent from
Figure~\ref{fig:Dikpati_datasets}, there are no big differences between
the results for the various datasets. In all cases, the drop from cycle
19 to cycle 20 is predicted well while the prediction for cycle 21 is
much too low. The correlation coefficients with the activity maxima of
the next cycle are $r=0.87$ for the dataset of
\citet{Balmaceda:etal:2005} and $r=0.73$ for the sunspot numbers, to be
compared with $r=0.90$ for the reference case.  The corresponding values
for cycle-integrated quantities are 0.83, 0.80, and 0.92.

The satisfactory predictive performance of the model with input from
monthly sunspot numbers suggests that we can use this dataset to include
more cycles. The result for cycles 1--23 is shown in
Figure~\ref{fig:Dikpati_sunspot_number}. The correlation between the
maxima of the flux diffusing over the equator and the maxima of the
sunspot number of the subsequent cycle is $r=0.80$. The two curves are
maximally correlated for a forward time shift of the $\Phi$ curve by
6.8~yr.

\subsubsection*{The relevance of the declining phase}

In order to illustrate the importance of the activity level during the
declining phase, we consider as a predictor the average sunspot activity
three years before the minima of the (unstretched) historical
cycles. The result based on the sunspot areas for cycles 12--23 are
shown in Figure~\ref{fig:Precursor_Laura}, while sunspot numbers for
cycles 1--23 are considered in Figure~\ref{fig:Precursor_SSN}. In both
figures, the predictor is shown by the diamond-shaped symbols. In order
to illustrate the predictive skill, we have marked the corresponding
activity level (multiplied by a factor 3) by circles at the times of the
following activity maxima and connected them by dashed lines. The
general trends and the significant drops of activity after cycles 4, 11,
and 19 are reproduced in both cases. The corresponding correlation
coefficients are $r=0.84$ for Figure~\ref{fig:Precursor_Laura} and
$r=0.89$ for Figure~\ref{fig:Precursor_SSN}. The values for using the
activity level 2 years and 4 years before minimum are $r=0.89$ and
$r=0.83$, respectively, for the sunspot numbers of cycles 1--23.
Consequently, the predictive skill of this very simple precursor is
comparable to the corresponding cases using the predictor $\Phi$ from
the flux transport model.  Together with the considerations presented in
connection with Figure~\ref{fig:blowup} this indicates that the level of
activity in the declining phase of a cycle is underlying the predictive
skill of our flux transport model when the source term is specified
according to the DGG model.

\subsubsection*{Changes of the source model}

Since the existing sunspot data since 1874 do actually contain the areas
and coordinates of each individual observed sunspot group, one can use
this information to make the source term more realistic. We have done so
in two steps, both based on the cross-calibrated sunspot group data set
provided by \citet{Balmaceda:etal:2005}.

1) In the first step, we change from the schematically prescribed fixed
   range and drift rate of the emergence latitudes to the average
   observed actual latitudes of the sunspot groups. This is done in the
   following way (cf. Figure~\ref{fig:data_emerg_lat}). We average
   the latitudes of the individual sunspot groups that appeared within
   one month, weighted by group area. Since a sunspot group typically
   appears more than once in the data, we consider a group only at the
   day of its maximum area in order to avoid multiple counting. When
   cycles overlap around solar minimum, sunspot groups near the equator
   are attributed to the old cycle while those appearing in higher
   latitudes are considered to be part of the new cycle (red and blue
   dotted areas in Figure~\ref{fig:data_emerg_lat}). For each cycle,
   the resulting monthly averages of the emergence latitudes (green
   curves in Figure~\ref{fig:data_emerg_lat}), are fitted to a
   parabola (black lines).  This procedure leads to a observationally
   based time-latitude profiles of flux emergence for each cycle, also
   allowing for the overlapping of cycles. Inspection of these parabolic
   profiles in Figure~\ref{fig:data_emerg_lat} shows that, for most
   cycles, they deviate strongly from the linear profiles between
   $35\deg$ and $5\deg$ assumed in the reference case. In fact, they are
   better represented by a linear profile between $25\deg$ and $5\deg$.
   
   We have run the flux transport model with these emergence profiles,
   considering separately the sunspot areas for each cycle in the source
   amplitudes, so that overlapping cycles are properly accounted for.
   Also, the tilt-angle factor $\sin\lambda$ has been included in the
   source amplitudes. All other parameters are as in the reference case
   discussed in
   Sect.~\ref{subsec:Dikpati}. Figure~\ref{fig:predict_emerg_lat} shows
   the resulting time evolution of the predictor $\Phi$ together with
   the observed (total) sunspot areas. Although the general trend of the
   cycle amplitudes is roughly reproduced, the value of $r=0.75$ for the
   correlation coefficient between the maxima shows that the predictive
   skill is significantly diminished with respect to the reference case.
   If we avoid the parabolic fit and directly use the monthly weighted
   averages (green curve in Figure~\ref{fig:data_emerg_lat}), the
   correlation coefficient drops to $r=0.43$. Likewise, omitting the
   monthly averages of the emergence latitudes and using a area-weighted
   direct parabolic fit through the emergence latitudes of the
   individual sunspot groups (red and blue dots in
   Figure~\ref{fig:data_emerg_lat}), yields $r=0.45$. In both cases, the
   predictive skill is almost completely lost.

   This leaves us with the surprising result that the predictive skill
   is strongly dimished or even almost lost when more detailed
   observational data are used for prescribing the source in the flux
   transport model. Moreover, the skill depends critically on the way
   that the emergence latitudes of the individual sunspot groups are
   averaged. These results become less puzzling when we consider in
   Figure~\ref{fig:data_emerg_lat} the broad distribution of the actual
   emergence latitudes for any given time: since the contribution to the
   predictor $\Phi$ depends sensitively on the emergence latitude, in the
   sense that near-equator emergence is much more strongly weighted,
   variations in the averaging procedure that lead to small changes 
   in the low-latitude part of the averaged emergence latitudes used in
   the flux-transport model can have a strong impact on the predictor
   quantity, $\Phi$. 

2) We can avoid the sensitivity of the results with respect to the
   averaging procedure of the emergence latitude by directly considering
   the contribution to the source term of each individual sunspot group,
   so that no averaging is required. Each sunspot group in the data is
   identified with a bipolar magnetic region that is introduced into the
   flux transport simulation at its recorded latitude. The magnetic flux
   content of the region is
%  proportional to its area \citep{Schrijve:Harvey:1994}, which, in
%  turn, is 
   assumed to be proportional to its maximum sunspot area. The
   orientation of the magnetic polarities follows Hale's laws and
   thus alternates from cycle to cycle. Taking the
   latitude separation of the two polarities of the region to be small
   compared to their diameter, which itself is assumed to be
   proportional to the square root of the sunspot area, we choose for
   the FWHM of the two Gaussians describing the radial field source
\begin{equation}
{\rm FWHM}= \left( {A_{\rm s}}\over{A_{\rm s,max}} \right) ^{1/2}
            \times 5 \dgr\,,
\label{eq:FWHM}
\end{equation}
where $A_{\rm s}$ is the area of the sunspot group and $A_{\rm s,max}$
the value for the largest sunspot groups in the dataset. For these
groups, we assume a latitude extension of $5\degr$, roughly
corresponding to half the size of the largest groups. It turns out that
the the results are rather insensitive to the exact value of this
parameter (see below). To account for the latitude dependence of the
tilt angle \citep{Howard:1991}, the amplitude of the corresponding
source term is multiplied by a factor $\sin\lambda$.

\begin{figure}[ht!]
\centering
\resizebox{1.0\hsize}{!}{\includegraphics[angle=-90]
          {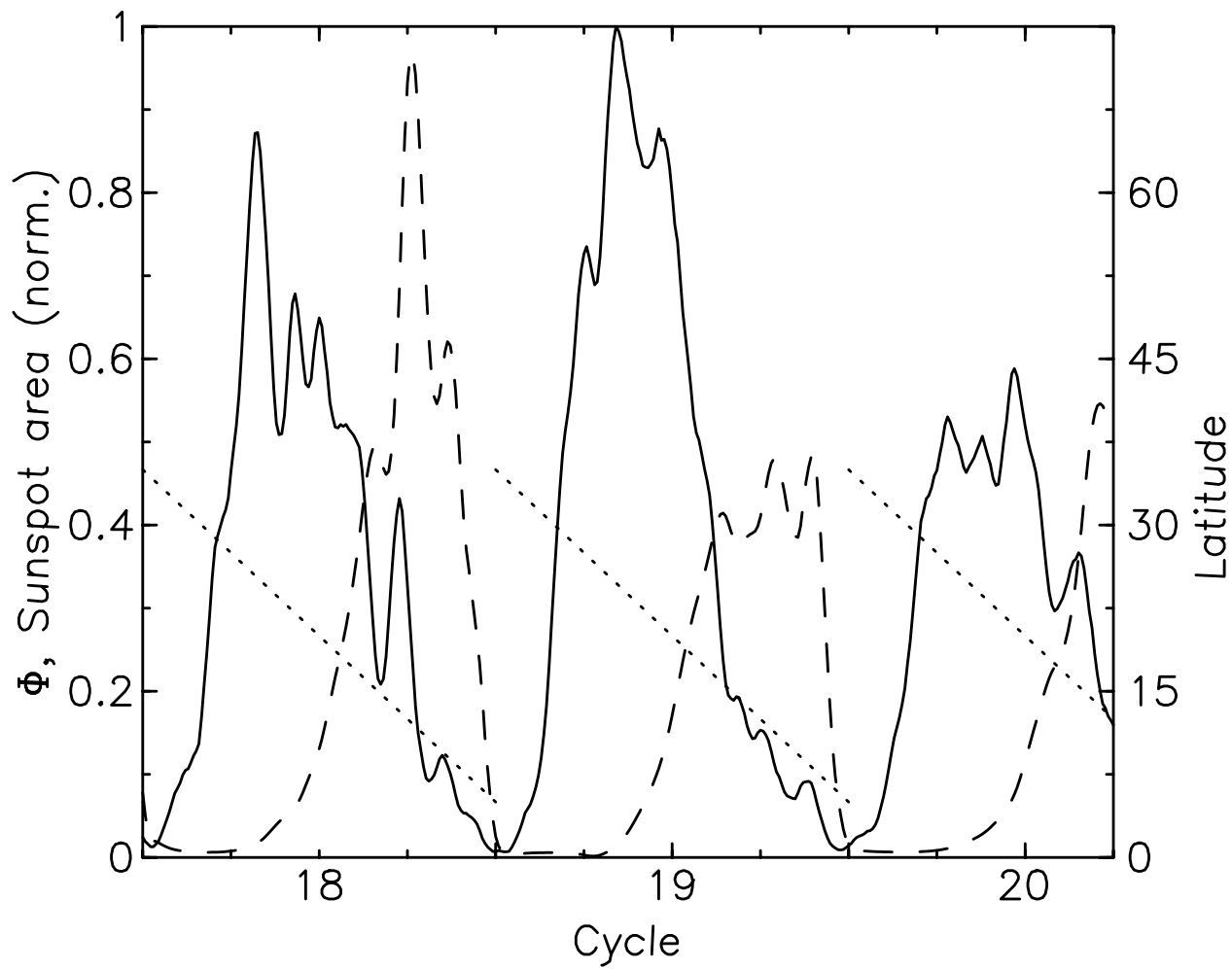}}
\caption{Normalized observed sunspot area (solid) and calculated
  magnetic flux diffusing over the equator ($\Phi$, dashed) for cycles
  18 to 20, together with the latitude progression of the centroid of
  the source term representing flux emergence in the model (dashed,
  scale to the right). The $\Phi$ curve peaks during the decling phases
  of the cycles, a few years before the minimum epochs. Their amplitude
  is determined by the level of low-latitude sunspot activity 
  (flux emergence) during this phase.}
\label{fig:blowup}
\end{figure}

\begin{figure}[ht!]
\centering
\resizebox{1.0\hsize}{!}{\includegraphics[angle=-90]
          {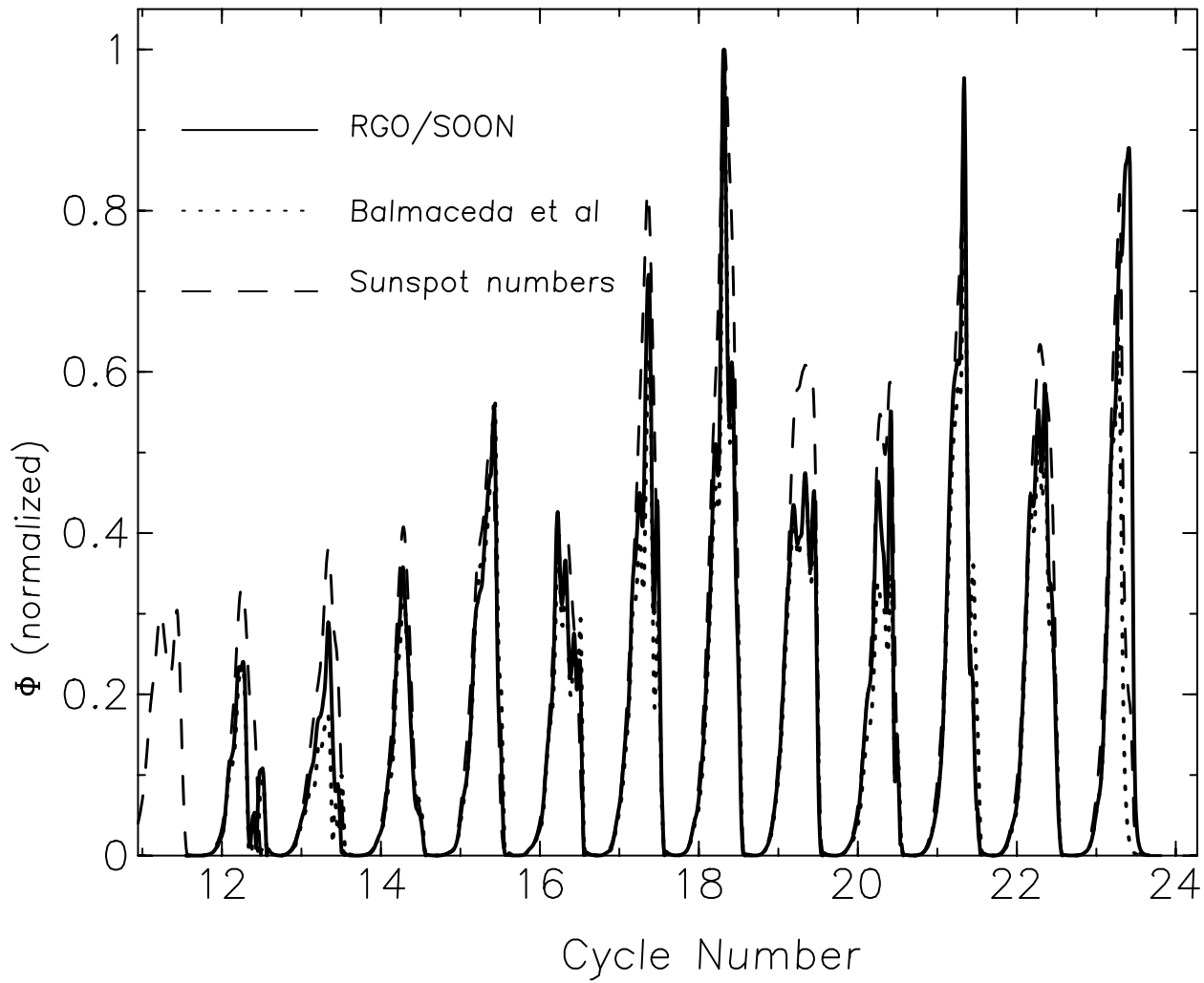}}
\caption{Magnetic flux diffusing over the equator per unit time
$(\Phi)$, determined from the surface flux transport model with input
from three different data sets.  Solid line: RGO/SOON data according to
D. Hathaway. Dotted line: cross-calibrated RGO, Russian and SOON data
according to \citet{Balmaceda:etal:2005}. Dashed line: monthly sunspot
numbers.  All curves are normalized to their respective maximum values.
The parameters and procedures in the flux transport simulation are
identical to those used for Figure~\ref{fig:Dikpati_reference}. The
correlation coefficients between the maxima of $\Phi$ and the activity
maxima of the subsequent cycles are 0.90 (RGO/SOON), 0.87
\citep{Balmaceda:etal:2005}, and 0.73 (sunspot numbers), respectively.}
\label{fig:Dikpati_datasets}
\end{figure}

\begin{figure*}[ht!]
\centering
\resizebox{1.0\hsize}{!}{\includegraphics[angle=-90]{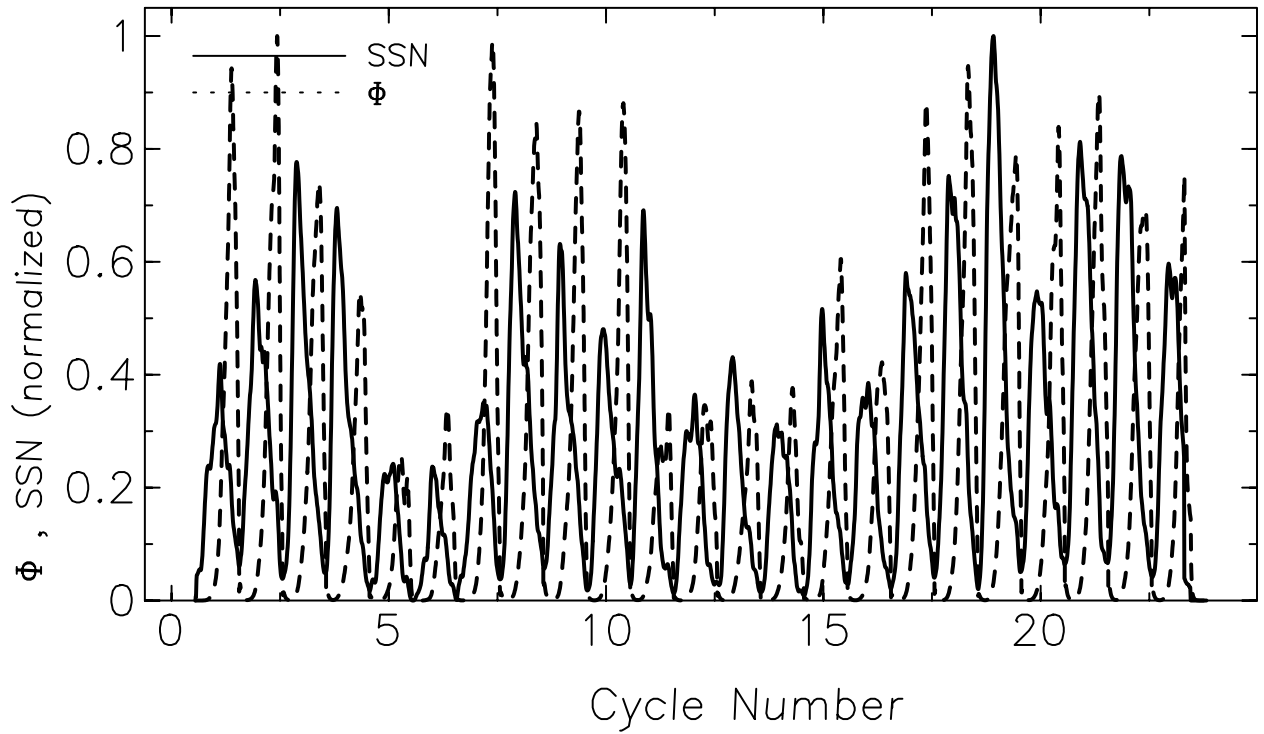}}
\caption{Magnetic flux diffusing over the equator ($\Phi$, dashed curve)
  from the surface flux transport model with input source based upon
  monthly sunspot numbers for cycles $1-23$ (SSN, solid line). Both
  curves are normalized to their respective maximum values. The
  correlation coefficient between the maxima of $\Phi$ and the maximum
  sunspot numbers of the subsequent cycles is $r=0.80$. The
  cross-correlation of both curves reaches a maximum for a forward shift
  of the $\Phi$ curve by 6.75~years.}
\label{fig:Dikpati_sunspot_number}
\end{figure*}

\begin{figure}[ht!]
\centering
\resizebox{1.0\hsize}{!}{\includegraphics[angle=-90]{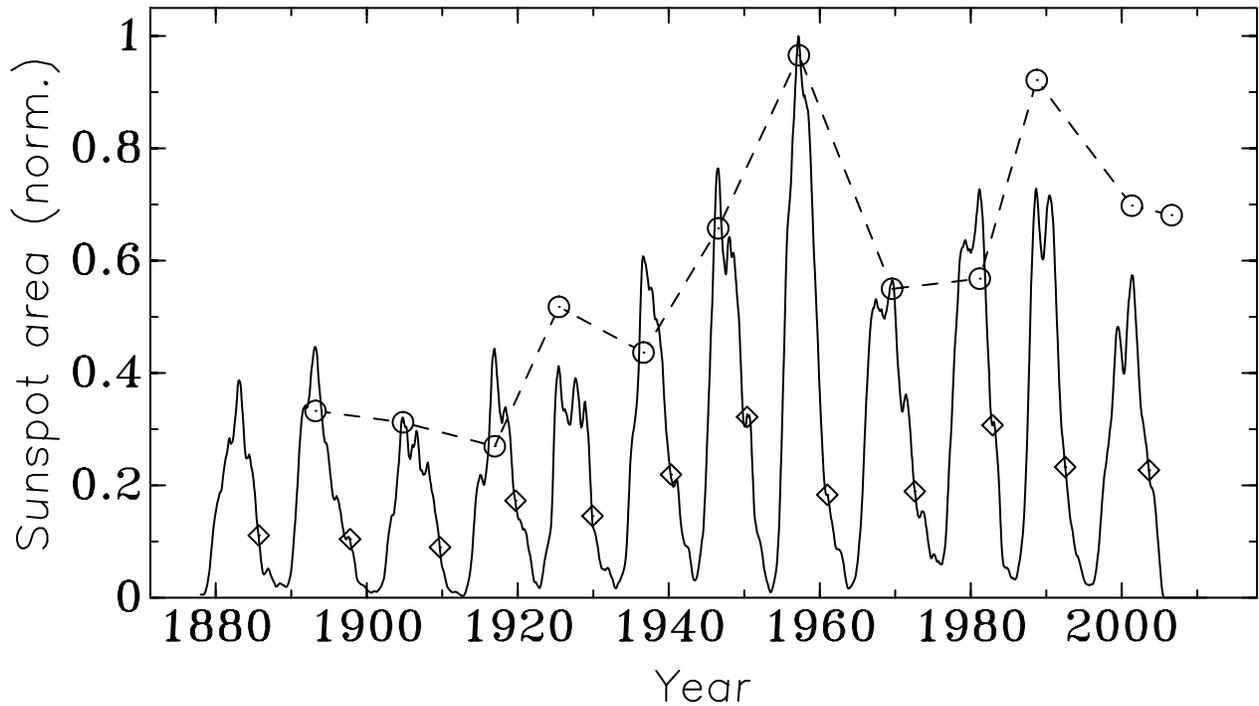}}
\caption{The activity level three years before sunspot minimum as a
  precursor for the amplitude of the subsequent cycle. The solid curve
  shows the normalized observed sunspot areas for cycles 12--23. The
  precursor is indicated by diamonds. In order to illustrate the
  predictive skill, open circles (connected by the dashed lines)
  representing 3 times the predictor value have been placed near the
  maximum epoch of the following cycles. The correlation coefficient
  between the predictor and the maximum sunspot area of the subsequent
  cycle is $r=0.84$.}
\label{fig:Precursor_Laura}
\end{figure}

\begin{figure}[ht!]
\centering
\resizebox{1.0\hsize}{!}{\includegraphics[angle=-90]{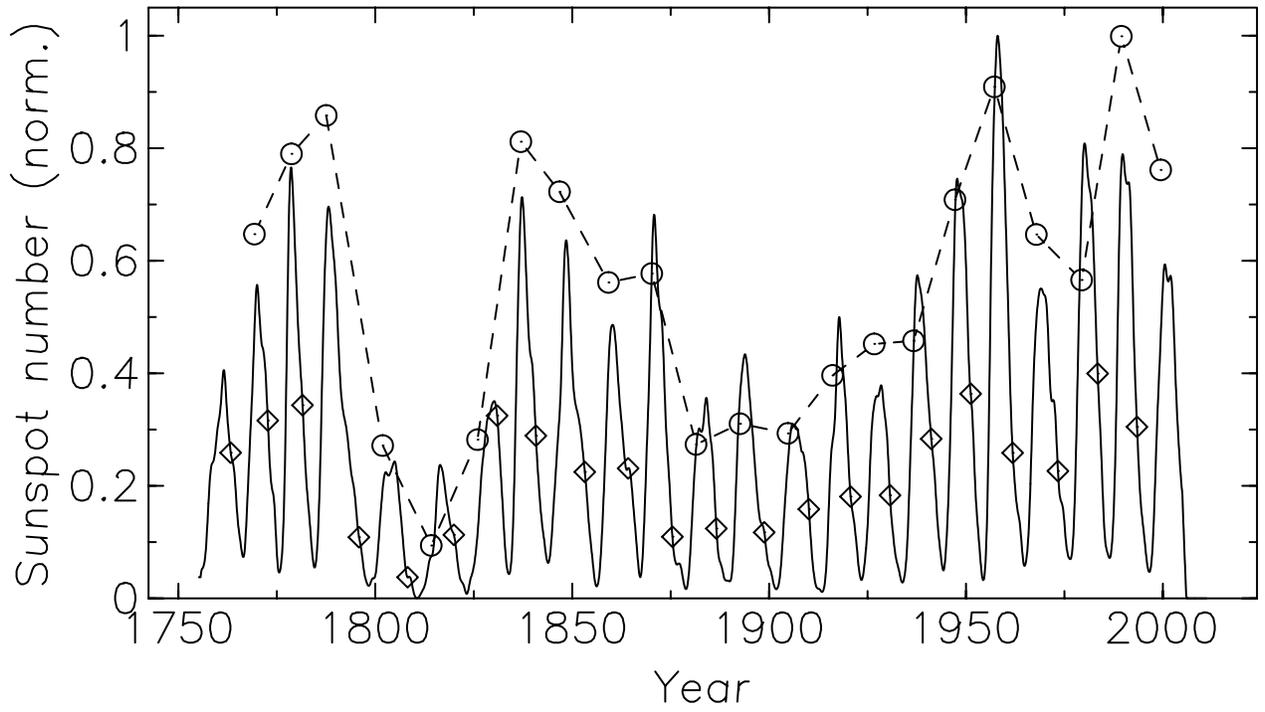}}
\caption{Same as Figure~\ref{fig:Precursor_Laura} with the sunspot
  numbers for cycles 1--23. The correlation coefficient is $r=0.89$.}
\label{fig:Precursor_SSN}
\end{figure}

\begin{figure}[ht!]
\centering
\resizebox{1.0\hsize}{!}{\includegraphics[angle=-90]{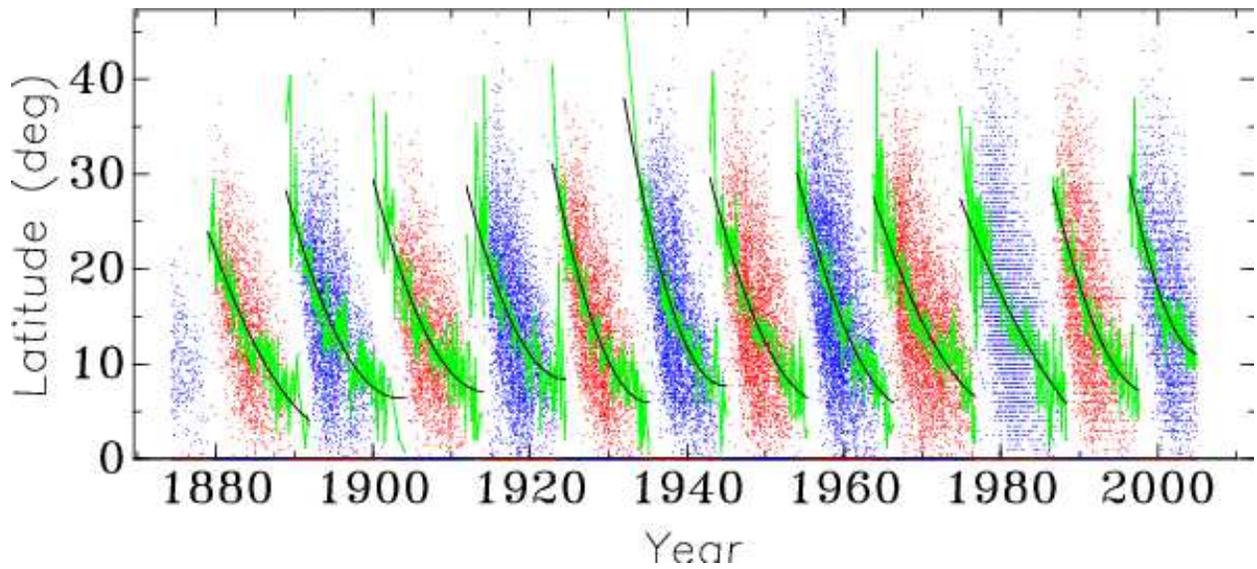}}
\caption{Emergence latitudes of sunspot groups for cycles 12--23. The
  blue and red dots indicate the individual sunspot groups as determined
  from the cross-calibrated data set of \citet{Balmaceda:etal:2005}.
  The green curves show monthly averages, weighted with the
  corresponding group areas, and the black lines are parabolic fits
  of the green curves. The fit curves are only drawn for the
  time intervals during which sunspots of the corresponding cycle have
  been actually observed. }
\label{fig:data_emerg_lat}
\end{figure}

\begin{figure}[ht!]
\centering
\resizebox{1.0\hsize}{!}{\includegraphics[angle=-90]{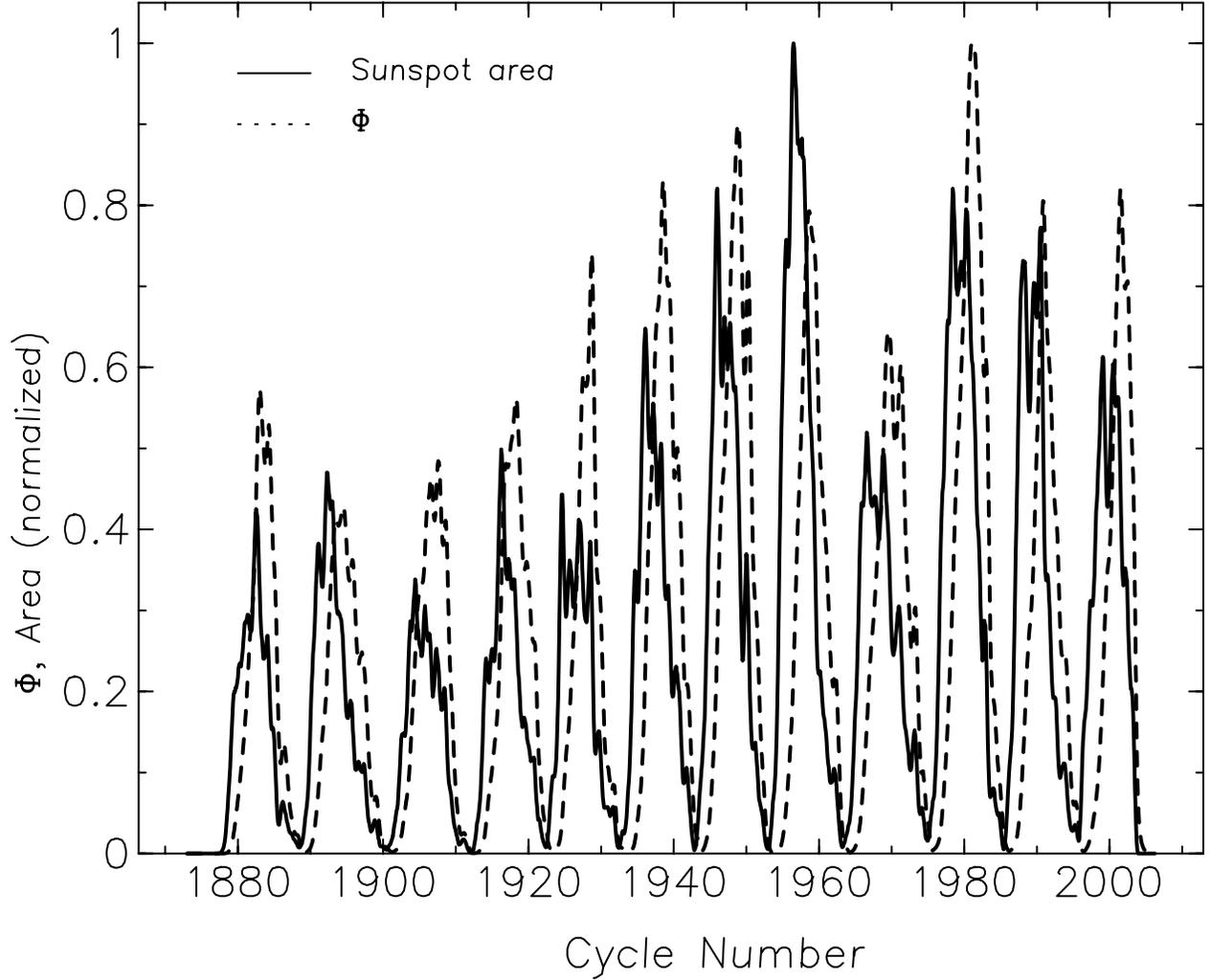}}
\caption{Comparison between the predictor quantity $\Phi$, the flux
  diffusing over the equator (dashed curve), and the observed sunspot
  area for cycles 12--23. $\Phi$ has been calculated with the
  flux-transport model with source input governed by the parabolic
  profiles of emergence latitudes shown in
  Figure~\ref{fig:data_emerg_lat}. The correlation coefficient between
  the maxima of the $\Phi$ and the maxima of the sunspot area of the
  subsequent cycle is $r=0.75$.  }
\label{fig:predict_emerg_lat}
\end{figure}

\begin{figure}[ht!]
\centering
\resizebox{1.0\hsize}{!}{\includegraphics[angle=-90]{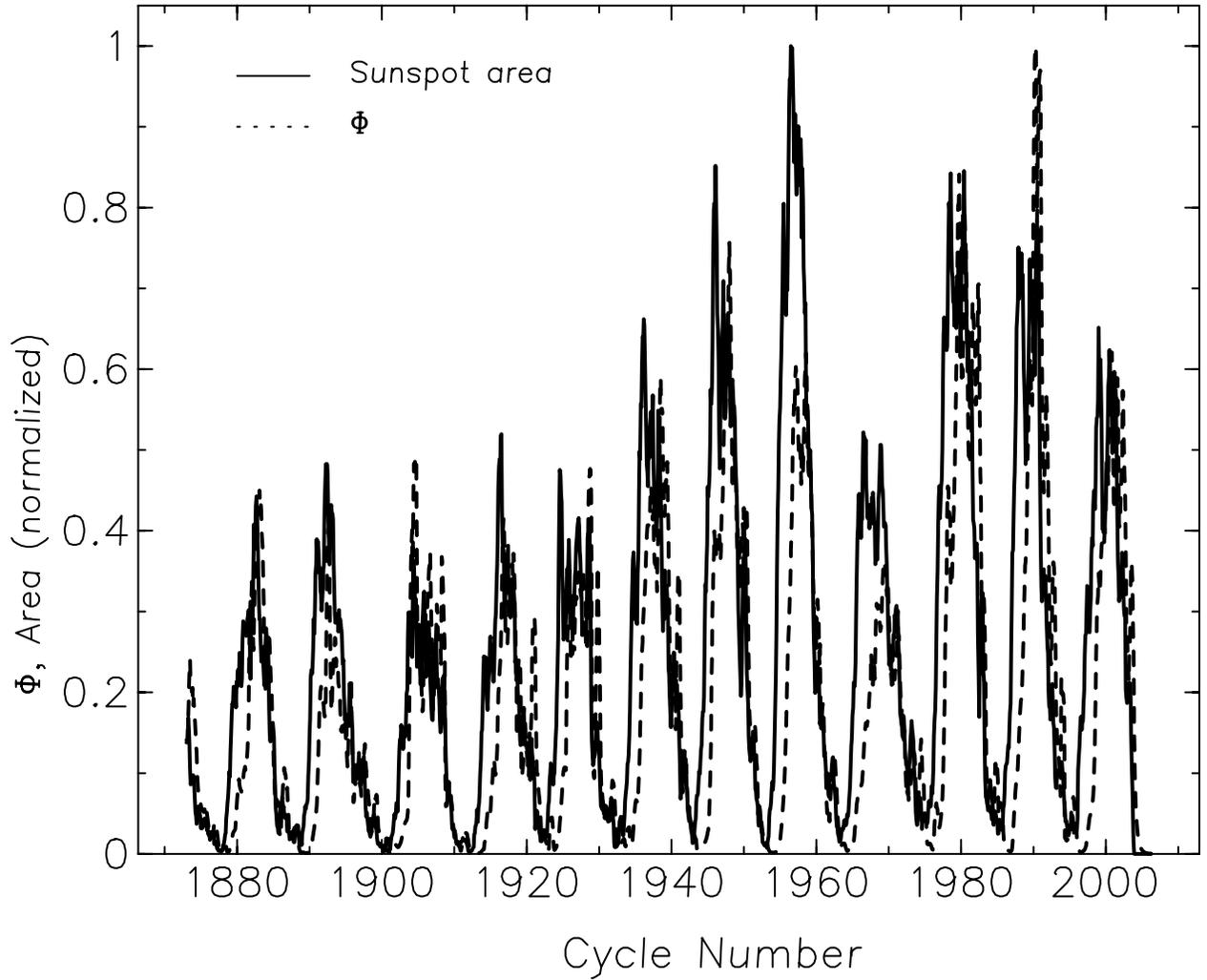}}
\caption{Same as Figure~\ref{fig:predict_emerg_lat} for a source term
  based upon the emergence of each individual sunspot group in the
  dataset of \citet{Balmaceda:etal:2005}.  The
  correlation coefficient between the maxima of the $\Phi$ and the
  maxima of the sunspot area of the subsequent cycle is $r=0.33$. }
\label{fig:actual_emerg_lat}
\end{figure}

Figure~\ref{fig:actual_emerg_lat} shows the prediction results. It is
obvious that the predictor, $\Phi$, now mainly reflects the strength of
the ongoing cycle and thus provides very low predictive
capability. Accordingly, the correlation coefficient between the maxima
of $\Phi$ and the sunspot area of the subsequent cycle is low: $r=0.33$
(the corresponding values for maximum latitude extensions of $10\deg$
and $2.5\deg$ are $r=0.27$ and $r=0.34$, respectively.  The origin of
this result is in the high sensitivity of $\Phi$ to bipolar regions
emerging in low latitudes (cf. Figure~\ref{fig:blowup}). Although the
tilt angle goes to zero, $\Phi$ is still dominated by these low-latitude
emergences. Since we have a broad distribution of emergence latitudes at
any given time, there is nearly always some flux emergence in low
latitudes (cf. Figure~\ref{fig:data_emerg_lat}. Because the amount of
this low-latitude flux emergence is mainly determined by the overall
strength of the {\it ongoing} cycle, the quantity $\Phi$ is no longer
dominated by the late phase as in the reference case, which did not take
into account the broad latitude range of emerging flux. Consequently,
the predictor mainly reflects the ongoing cycle, so that the predictive
skill of the model is a nearly completely lost. Note that this version
of the model makes the most direct use of the actual data.

A result pointing in this direction appears already in
our parameter study (see Table~\ref{tab:ccs}): when doubling the
latitudinal width of the source term (to 12 degrees FWHM of the
corresponding Gaussian), the correlation coefficient dropped from 0.90
to 0.66. When we impose a narrow latitudinal extension of the source
term, low-latitude emergence occurs exclusively in the declining phase
of a solar cycle, i.e., during a few years before sunspot minimum. The
more activity (flux emergence) during the this phase, the more flux
diffuses over the equator, and the higher is the amplitude of our
predictor (cf. Figure~\ref{fig:blowup}).

\section{What is the origin of the predictive skill?}
\label{sec:origin}

We have seen that our predictor $\Phi$ determined with the flux
transport model provides reasonable predictive skill in the case of a
narrow latitudinal width and a prescribed fixed latitude range of the
source term and we have shown that this is related to the activity level
in the declining phase of the cycle. However, the predictive skill of
$\Phi$ is strongly dimished or even nearly completely lost when averages
of the actually observed observed emergence latitudes are considered or
if the individual emergence latitudes of sunspot groups are used
directly. This result casts serious doubt upon connecting the predictive
skill in the standard case with the Babcock-Leighton dynamo scheme and
with the dipole strength of the surface field. But how else can we
understand the unquestionable correlation between the activity level in
the declining phase of a cycle and the strength of the next cycle?

We suggest that in fact no direct physical link between the surface
manifestations of the old and the new cycle is required to understand
the predictive skill of the activity level in the declining phase. All
that is needed are two well-established properties of the sunspot
record: 1) the temporal overlapping of cycles with high-latitude spots
of the new cycle already appearing when the old cycle is still in
progression in low latitudes \citep[e.g.,][]{Harvey:1992a} and 2) the
relation between the rise time of a cycle towards its maximum and its
strength: stronger cycles rise faster towards sunspot maximum
\citep[known as the Waldmeier effect, e.g.][]{Waldmeier:1935}. When
considering latitude-integrated quantities like the sunspot number or
area, the combined effect of both properties leads to a systematic shift
of the minimum epochs between two cycles of different strength: the
minimum occurs earlier if the following cycle is stronger than the
preceding cycle and later for a weaker following cycle. This effect
explains the empirical statistical relationship between cycle length and
amplitude: strong cycles tend to be preceded by short cycles
\citep[e.g.,][]{Hathaway:etal:1999, Hathaway:etal:2002,
Solanki:etal:2002b}.

Figure~\ref{fig:idea} illustrates how the overlapping of cycles with
amplitude-dependent rise time leads to a predictive skill of the
activity level during the declining phase of the preceding cycle. In
this figure we have considered time profiles of the activity cycles
according to a prescribed functional form
\citep{Li:1999,Hathaway:etal:1994}, which reproduces both the rise and
decay parts of a cycle, including the Waldmeier effect:
\begin{equation}
  f(t) = \frac{a(t-t_0)^3}{e^{(t-t_0)/b}-c}\,,
\label{eq:Li_f}
\end{equation}
where $t_0$ denotes the starting time of the cycle. The parameter
$b=1.128\,$yr and the relation between  $c$ and  $a$
of the form
\begin{equation}
  c = -1.104\,10^{-4}\,a^2 + 0.24666\,a - 123.593
\label{eq:Li_c}
\end{equation}
have been determined by \citet{Li:1999} on the basis of sunspot area
data since 1876. By fitting the two parameters, $t_0$ and $a$, the
cycles 12--22 are well reproduced by the functional form given by
equation (\ref{eq:Li_f}). The left panel of Figure~\ref{fig:idea} shows
the summed activity levels of a cycle followed by a stronger cycle
(solid curve) or by a weaker cycle (dashed curve), respectively, the
follower cycle starting 11 years after the first cycle in both cases.
The individual cycles profiles are indicated by the dotted lines. The
faster rise of the stronger following cycle leads to a earlier `sunspot
minimum' than in the case of a weaker follower (marked in the figure by
`M1' and `M2', respectively), the time shift being about one year.
Since observationally a sunspot cycle is defined as the time between
adjacent minima, the activity in the declining phase of the first cycle,
(i.e., in a fixed time interval relative to the respective solar minimum
epoch) is considerably larger when the follower cycle is stronger than
when it it weaker \citep{Hathaway:etal:1999}. In our example, for
instance, the activity level three years before the respective minimum
is about a factor of two larger for the stronger following cycle than
for the weaker follower (marked in the figure by `P1' and `P2'),
corresponding to the relative strength of the two follower
cycles. Consequently, the activity level at time P1/P2 can be used as a
predictor for the amplitude of the following cycles, but it can only be
determined after the epoch of the minimum is known; before, only upper
limits can be given.

The left panel of Fig.~\ref{fig:idea} shows that the above defined
predictor P1/P2 (activity level three years before cycle minimum) is
only weakly affected by the strength of the first cycle. This results
from the fact that the declining phase of a cycle largely represents an
exponential decay, irrespective of the cycle strength.
Fig.~\ref{fig:idea} also suggests that the shift of sunspot minima
depending on the amplitudes of overlapping cycles leads to a correlation
of the activity level at sunspot minimum with the amplitude of the
following cycle. In fact, \citet{Hathaway:etal:2002} find a correlation
coefficient of $r=0.72$ between these quantities for the historical
sunspot record.

As a consequence, simply two well-known properties of the solar cycles,
the Waldmeier effect and the overlapping, explain the predictive skill
of the activity level in the declining phase of a cycle. This explanation also
carries over to the skill of the our predictor from the flux transport
model, the amount of flux diffusing over the equator: assuming for the
source term a linear profile of emergence latitude versus time between
two sunspot minima and an amplitude proportional to the instantaneous
activity level, leads to a stronger source in low latitudes (declining
phase of the cycle) for an earlier minimum, which, in turn, results from
a stronger following cycle. Through the dependence  of the sunspot
minimum shift on the amplitude of the next cycle,
information about the strength of the following cycle propagates into
our predictor, completely independent of whether there is a physical
connection between surface manifestations of the cycles or not. 
No further `memory' of the system is required.

We demonstrate the effect of the minimum shift on the prediction with a
precursor method by considering series of overlapping synthetic cycles
with profiles according to equation~(\ref{eq:Li_f}). The cycle
amplitudes form a random sequence with a flat probability density
between 500 and 3000 microhemispheres for the maximum sunspot area, so
that the strengths of subsequent cycles are completely independent of
each other. The cycles start at regular intervals of $10.75\,$yr;
because of their overlapping, the cycle length defined by the time
between subsequent activity minima varies. As an example,
Figure~\ref{fig:synthetic} shows the analog to
Figure~\ref{fig:Precursor_SSN} for one realization of 23 random
synthetic cycles. Statistics of this synthetic series, such as the
correlation of the minima with the amplitudes of the subsequent cycle
($r=0.60$) and the correlation between the length of a cycle and the
amplitude of the next cycle ($r=-0.82$) are roughly consistent with the
properties of the actual sunspot record \citep[exhibiting values of 0.72
and $-0.69$, respectively, see][]{Hathaway:etal:1999}, which is
sufficient for the purpose of illustration. Using the activity level 3
years before the activity minima as predictor, we find about the same
skill for the synthetic random series (correlation coefficient $r=0.84$)
as for the actual data ($r=0.89$, see Figure~\ref{fig:Precursor_SSN}).
This demonstrates that the amplitude-dependent overlapping of cycles
affects the timing of the activity minima in such a way that information
from cycle $n+1$ can be picked up by considering the activity in the
declining phase of cycle $n$, {\em but only after the epoch of the
minimum is known}. As we have seen, this effect allows us to `predict' a
random sequence of cycle amplitudes without any relation between
subsequent cycles apart from overlapping.

For 1000 random series of 8 cycles each, Figure~\ref{fig:synthetic_ccs}
shows the cumulative probability distribution for the correlation
coefficient between the activity level three years before minimum and
the subsequent maximum. The median value corresponds to $r=0.83$, i.e.,
in 50\% of the cases the correlation coefficient is larger than this
value; it is larger than 0.95 in 5\% of the cases. These results with
random sequences of cycle amplitudes indicate that correlation
coefficients in the range of 0.8--0.9, which are regularly found with
precursor methods, may be easily be explained by the minimum shift
effect alone and require no further physical connection of the surface
quantities between subsequent cycles. This does not mean that such a
connection may not exist, only that correlations of that size do not
compel us to assume such a connection.

\section{Discussion and conclusions}
\label{sec:discussion}

The DDG model is able to reproduce the amplitudes of cycles 16--23 with
impressively large correlation coefficients (up to $r\simeq 0.99$) and
we have seen that our very much simpler model exhibits somewhat less (up
to $r\simeq 0.95$), but still considerable predictive skill, provided
that we closely follow the DDG treatment of the surface source term.  In
fact, for a Babcock-Leighton type flux-transport dynamo, our predictor,
the flux diffusing over the equator determines the strength of the
reversed global dipole field from which differential rotation generates
the toroidal flux for the next activity cycle. Therefore, it is tempting
to consider these correlations as indicating that such a type of dynamo
model in fact represents the engine underlying the solar activity cycle.

However, at least in the case of our model, the predictive skill is
nearly completely lost when we use more of the available observational
information about flux emergence in the photosphere, particularly
concerning the emergence latitudes of bipolar magnetic regions. If
we allow for the overlapping of cycles and use the full
information provided by the butterfly diagram, the predictor largely
reflects the strength of the ongoing cycle and exhibits almost no
relation to the amplitude of the next cycle.

We have shown that this puzzling result, i.e., the schematic source
providing predictive skill while the source based on actual data showing
none, can be understood by the strong dependence of the schematic source
on the epochs of the sunspot minima. The overlapping of cycles whose
rise time is anticorrelated their amplitude (the Waldmeier effect)
naturally leads to a time shift of the minima that is strongly related
to the strength of the following cycle, thus affecting the strength of
the schematic source for low-latitude emergence in the declining phase of a
cycle. We have demonstrated the importance of this effect by showing
that such propagation of amplitude information from the next cycle into
the minimum epochs allows us to `predict' synthetic cycles with random
amplitudes with about the same skill as for the actual sequence of
sunspot cycles.

\begin{figure*}[ht!]
\centering
\resizebox{1.0\hsize}{!}{\includegraphics[angle=-90]{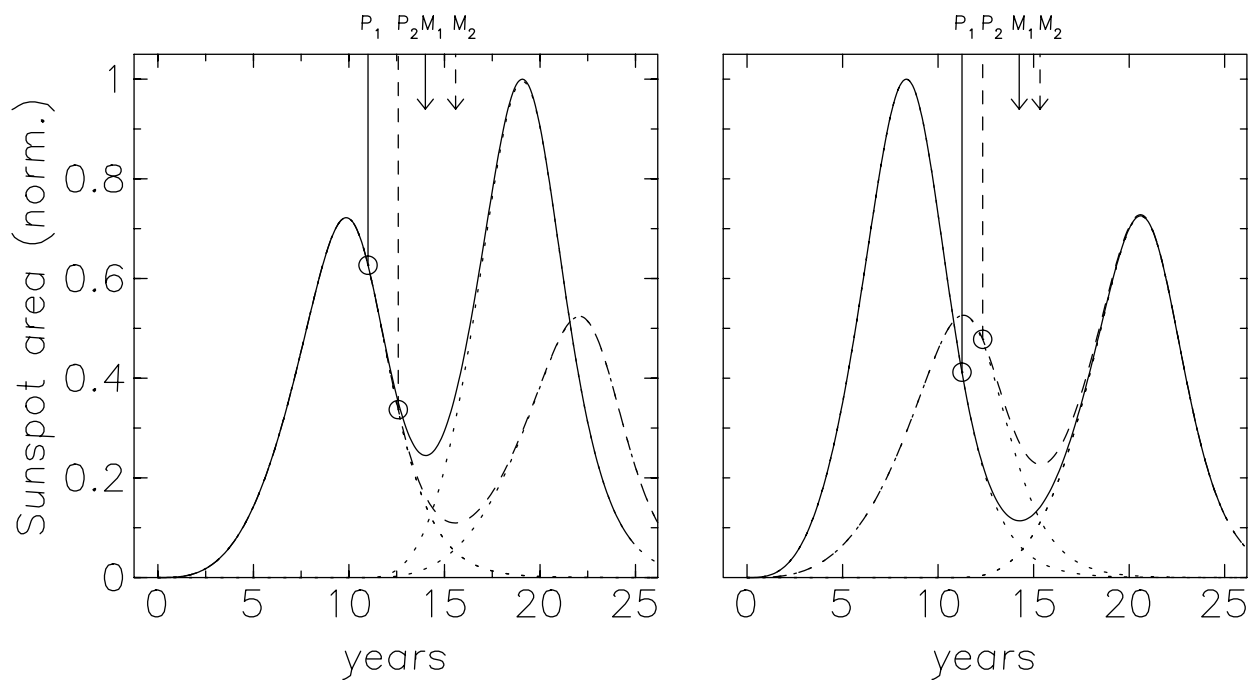}}
\caption{Schematic illustration of the amplitude-dependent shift of the
  minima between overlapping sunspot cycles and its influence on
  precursor quantities. Left panel: effect of the following cycle. A
  stronger follower cycle (solid curve) with a shorter rise time leads
  to an earlier minimum (M1) and a higher predictor (P1) than a weaker
  subsequent cycle with a longer rise time (dashed curve, minimum M2,
  predictor P2). Right panel: effect of the preceding cycle. Since the
  cycle asymmetry mainly affects the rising phase, the influence of the
  strength of the preceding cycle on the minimum epoch and on the
  predictor is much smaller.}
\label{fig:idea}
\end{figure*}

\begin{figure}[ht!]
\centering
\resizebox{1.0\hsize}{!}{\includegraphics[angle=-90]{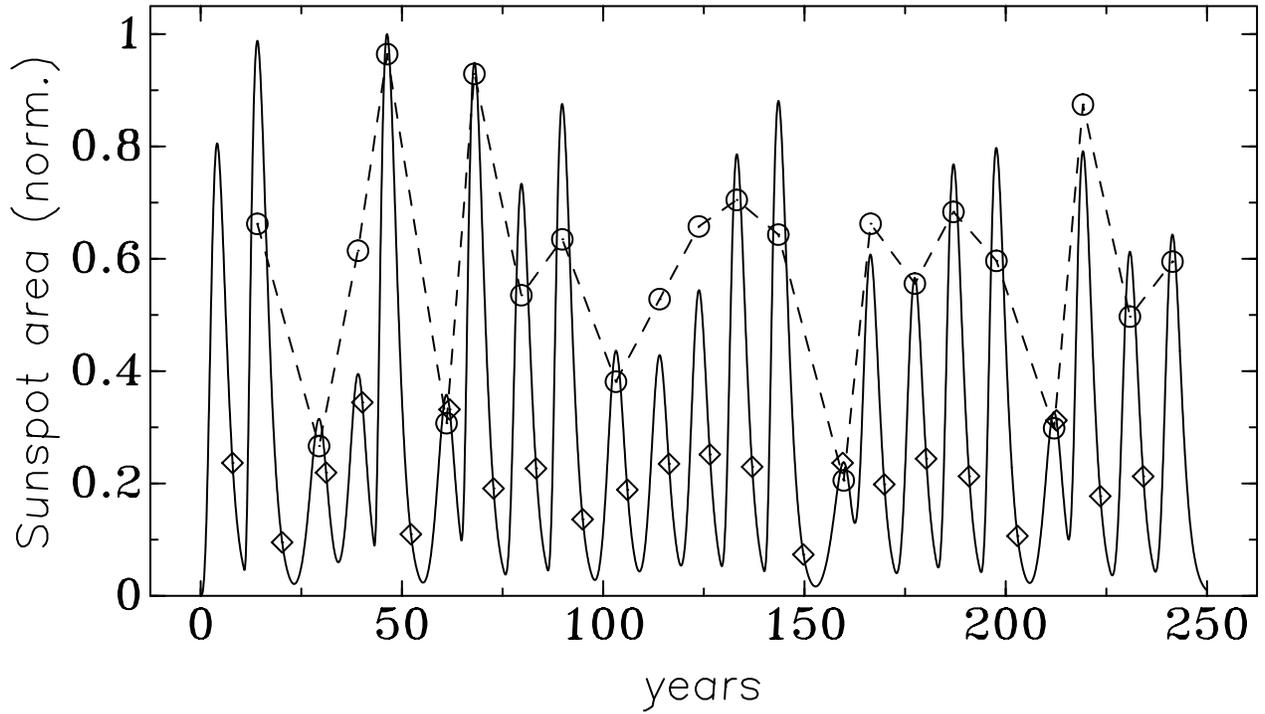}}
\caption{Comparison of the activity three years before minimum with the
  maximum of the subsequent cycle for a series of 23 synthetic solar
  cycles with a random distribution of maximum sunspot area between
  500 and 3000 millionths of a solar hemisphere. The quality of the
  `prediction' of the random amplitudes (correlation coefficient
  $r=0.84$) is comparable to that based on the real data (see
  Figure~\ref{fig:Precursor_SSN} with $r=0.89$).}
\label{fig:synthetic}
\end{figure}

\begin{figure}[ht!]
\centering
\resizebox{1.0\hsize}{!}{\includegraphics[angle=-90]{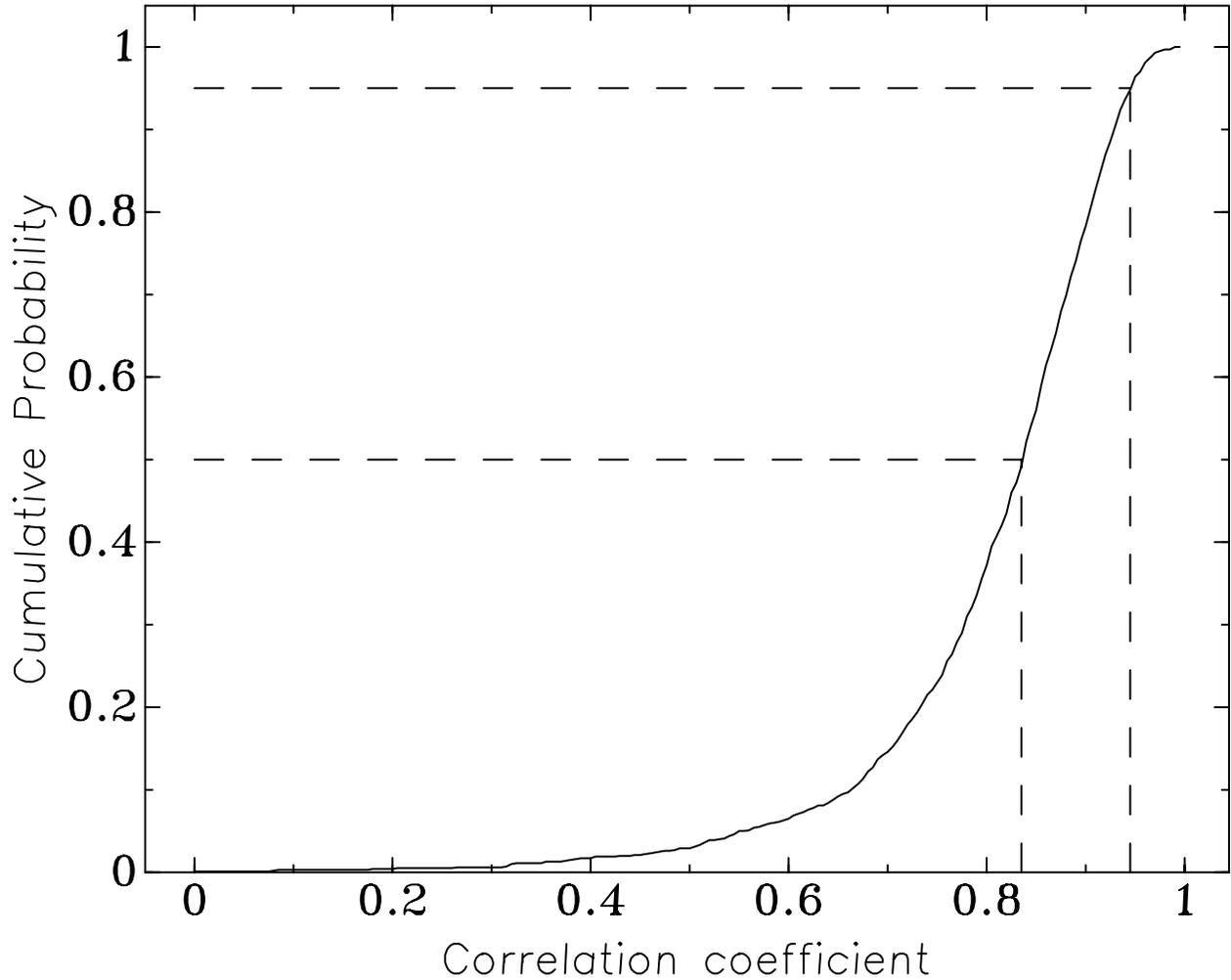}}
\caption{Cumulative probability distribution function for the
  correlation coefficients between the maximum sunspot number of a cycle
  and the sunspot number three years before the preceding minimum for
  1000 synthetic datasets of 8 cycles each. The cycles have random
  amplitudes between 500 and 3000 millionths of a solar hemisphere. The
  correlation coefficient is larger than 0.83 in 50\% of the cases and
  exceeds 0.95 in 5\% of the cases.}
\label{fig:synthetic_ccs}
\end{figure}

The amplitude-dependent minimum shift explains the results obtained with
our flux transport model. Moreover, it could account for the (partial)
success of a number of precursor methods \citep[for a summary,
see][]{Hathaway:etal:1999}. This includes the correlation of the sunspot
maximum with the activity level during the preceding minimum and the
anticorrelation with the length of the preceding cycle
\citep{Wilson:etal:1998,Solanki:etal:2002b}, which directly result from
the amplitude-related shift of the minimum epoch (see Figure
\ref{fig:idea}). Likewise, the anticorrelation of the skewness of the
cycle profile with the amplitude of the subsequent cycle
\citep{Lantos:2006} can be understood by the minimum shift: a weak
follower cycle is preceded by a late minimum, so that the foregoing
cycle has a longer decay time and thus becomes more asymmetric, and vice
versa.

The amplitude-dependent minimum shift also explains the predictive skill
of some geomagnetic precursors, such as the minimum level of the {\em
aa} index \citep{Ohl:1966}. Other precursor methods are not so easily
reduced to the minimum effect. For instance, when considering cycles
12-22 in retrospect, the method of \citet{Thompson:1993} yields a
coefficient of 0.97 between the number geomagnetically disturbed days
($Ap>25$) during cycle $n$ with the sum of the sunspot number maxima of
cycle $n$ and $n+1$ \citep{Hathaway:etal:1999}. For cycle 23, however,
this method predicted a maximum (yearly) sunspot number of about 160,
more than 30\% too large. It is therefore unclear whether this method
has a real physical basis or if the good correlation in the past is just
fortuitous, the method being just one of a large number of other
possibilities and happened to provide a large correlation coefficient
for the existing data. Some methods depending on a ad-hoc recipes for
the selection or division of geomagnetic data during the declining phase
of the sunspot cycle into an `activity-related' and an `interplanetary'
component \citep{Legrand:Simon:1981,Feynman:1982} have also been shown
to provide high correlations for past cycles
\citep[e.g.,][]{Hathaway:Wilson:2006}.  They may capture early
manifestations of the (extended) new cycle in middle to high latitudes,
for instance, in the form of ephemeral active regions
\citep{Harvey:1992a}, but the physical connection remains to be
understood.  In any case, it appears to be the early influence of the
{\em new} cycle which makes a prediction possible.

In summary, we have shown that the predictive skill of many precursor
methods and that of our simple flux transport model with a schematic
source can be reduced to the time shift of solar minima depending on the
strength of the next cycle. Therefore, information about the next cycle
is available (in statistical sense, of course) at the time when the
solar minimum is clearly identified -- not always an easy and
straightforward task, see \citet{Harvey:White:1999}.  Concerning our
physical understanding of the solar cycle and the dynamo mechanism, the
crucial point is that the predictive skill of such precursors does
neither require nor imply a physical relation between surface
manifestations of subsequent cycles -- we have shown that such methods
can be applied to for random sequences of cycles with fully independent
amplitudes.

In any case, one should not be misled by high correlation coefficients
for reproducing the past because most methods can be adjusted once the
actual numbers are known.  At best, precursor methods could indicate a
trend for the next cycle, but we cannot expect to fully capture the
intrinsic variability of the solar dynamo process with such simple
recipes. Our work gives no indication that flux-transport models of the
surface field could improve the predictions, quite the contrary: our
model shows no predictive skill when the actually observed emergence
latitudes are used. There is no reason to assume that a more
sophisticated 2D surface flux transport model would lead to a different
result. The predictive skill arising in the case of a schematic
prescription of the emergence latitudes results from the shift of the
minimum epoch depending on the strength of the {\em next} cycle, i.e.,
on information from the cycle to be predicted. We can tune the
parameters of the model to obtain correlation coefficients with past
cycles of $r\simeq 0.95$, but much simpler precursors (like the
activity level 3 years before minimum) also reach values around 0.90.

While lacking the sub-surface transport and the generation of the
toroidal field, our flux-transport approach is conceptually similar to
the model of \citet{Dikpati:etal:2006} and
\citep{Dikpati:Gilman:2006}. We conjecture that the amplitude-dependent
minimum shift should have an impact on the predictive skill of their
model as well. This could be easily tested by replacing the schematic
latitude dependence of the source by the actual emergence latitudes for
each sunspot group, thus avoiding the minimum shift effect.

\acknowledgments {Extended discussions with Mausumi Dikpati and Peter
 Gilman about the DGG model are gratefully acknowledged. Laura Balmaceda
 kindly put to our disposal her cross-calibrated dataset of sunspot
 group areas.  Helpful comments by an anonymous referee led to a
 substantial improvement of the presentation in various parts of the
 paper.}

%\bibliographystyle{apj}
%\bibliography{apj-jour,papref}
%\bibliography{joushort,papref}
\bibliography{ms}

\clearpage

\end{document}